\documentclass{webofc}
\usepackage[varg]{txfonts}   

\usepackage{amssymb}
\usepackage{graphicx}
\usepackage{float}
\usepackage[table]{xcolor}
\usepackage{booktabs}
\usepackage{multicol}
\usepackage{multirow}
\usepackage{longtable}
\usepackage{hyperref}
\usepackage{amsmath}
\usepackage{CERNWUTcolors}

\makeatletter
\newcommand*\rel@kern[1]{\kern#1\dimexpr\macc@kerna}
\newcommand*\widebar[1]{%
  \begingroup
  \def\mathaccent##1##2{%
    \rel@kern{0.8}%
    \overline{\rel@kern{-0.8}\macc@nucleus\rel@kern{0.2}}%
    \rel@kern{-0.2}%
  }%
  \macc@depth\@ne
  \let\math@bgroup\@empty \let\math@egroup\macc@set@skewchar
  \mathsurround\z@ \frozen@everymath{\mathgroup\macc@group\relax}%
  \macc@set@skewchar\relax
  \let\mathaccentV\macc@nested@a
  \macc@nested@a\relax111{#1}%
  \endgroup
}
\makeatother

\newcommand{\pt}           {\ensuremath{p_{\rm T}}\xspace}

\newcommand{\nineH}        {$\sqrt{s}~=~0.9$~Te\kern-.1emV\xspace}
\newcommand{\seven}        {$\sqrt{s}~=~7$~Te\kern-.1emV\xspace}
\newcommand{\twoH}         {$\sqrt{s}~=~0.2$~Te\kern-.1emV\xspace}
\newcommand{\twosevensix}  {$\sqrt{s}~=~2.76$~Te\kern-.1emV\xspace}
\newcommand{\five}         {$\sqrt{s}~=~5.02$~Te\kern-.1emV\xspace}
\newcommand{\twosevensixnn}{$\sqrt{s_{\mathrm{NN}}}~=~2.76$~Te\kern-.1emV\xspace}
\newcommand{\fivenn}       {$\sqrt{s_{\mathrm{NN}}}~=~5.02$~Te\kern-.1emV\xspace}

\newcommand{\GeVc}         {Ge\kern-.1emV/$c$\xspace}
\newcommand{\MeVc}         {Me\kern-.1emV/$c$\xspace}
\newcommand{\TeV}          {Te\kern-.1emV\xspace}
\newcommand{\GeV}          {Ge\kern-.1emV\xspace}
\newcommand{\MeV}          {Me\kern-.1emV\xspace}
\newcommand{\GeVmass}      {Ge\kern-.2emV/$c^2$\xspace}
\newcommand{\MeVmass}      {Me\kern-.2emV/$c^2$\xspace}

\begin{document}
\title{Particle identification with machine learning in ALICE Run~3}

\author{\firstname{Maja} \lastname{Karwowska}\inst{1,2}\fnsep\thanks{\email{mkabus@cern.ch}} \and
        \firstname{Monika} \lastname{Jakubowska}\inst{3} \and
        \firstname{Łukasz} \lastname{Graczykowski}\inst{2} \and
        \firstname{Kamil} \lastname{Deja}\inst{4,5} \and
        \firstname{Miłosz} \lastname{Kasak}\inst{4}
}

\institute{CERN -- European Organization for Nuclear Research
\and
           Faculty of Physics, Warsaw University of Technology
\and
           Faculty of Electrical Engineering, Warsaw University of Technology
\and
           Faculty of Electronics and Information Technology, Warsaw University of Technology
\and
           IDEAS NCBR
          }

\abstract{%
The main focus of the ALICE experiment, quark--gluon plasma measurements, requires accurate particle identification (PID). The ALICE sub-detectors allow identifying particles over a broad momentum interval ranging from about 100 \MeVc up to 20 \GeVc.
However, a machine learning (ML) model can explore more detector information. During LHC Run 2, preliminary studies with Random Forests obtained much higher efficiencies and purities for selected particles than standard techniques.

For Run 3, we investigate Domain Adaptation Neural Networks that account for the discrepancies between the Monte Carlo simulations and the experimental data. Preliminary studies show that domain adaptation improves particle classification. Moreover, the solution is extended with Feature Set Embedding and attention to give the network more flexibility to train on data with various sets of detector signals. PID ML is already integrated with the ALICE Run 3 Analysis Framework. Preliminary results for the PID of selected particle species, including real-world analyzes, are discussed as well as the possible optimizations.
}

\maketitle

\section{Introduction}
\label{sec:intro}

ALICE (A Large Ion Collider Experiment)~\cite{ALICE:2008ngc} is one of the four big detectors located at the Large Hadron Collider (LHC) at CERN. ALICE studies the properties of quark--gluon plasma (QGP), a hot and dense state of matter, and the strong force that holds quarks together inside hadrons~\cite{ALICE:2022wpn}. Detailed analysis of QGP requires accurate particle identification (PID), i.e., the ability to discriminate between different particle species produced during the collision. High PID precision distinguishes ALICE from other LHC experiments and allows for selecting a subset of particles required for specific analysis.

\begin{figure}[h]
    \centering
    \includegraphics[width = .7\textwidth]{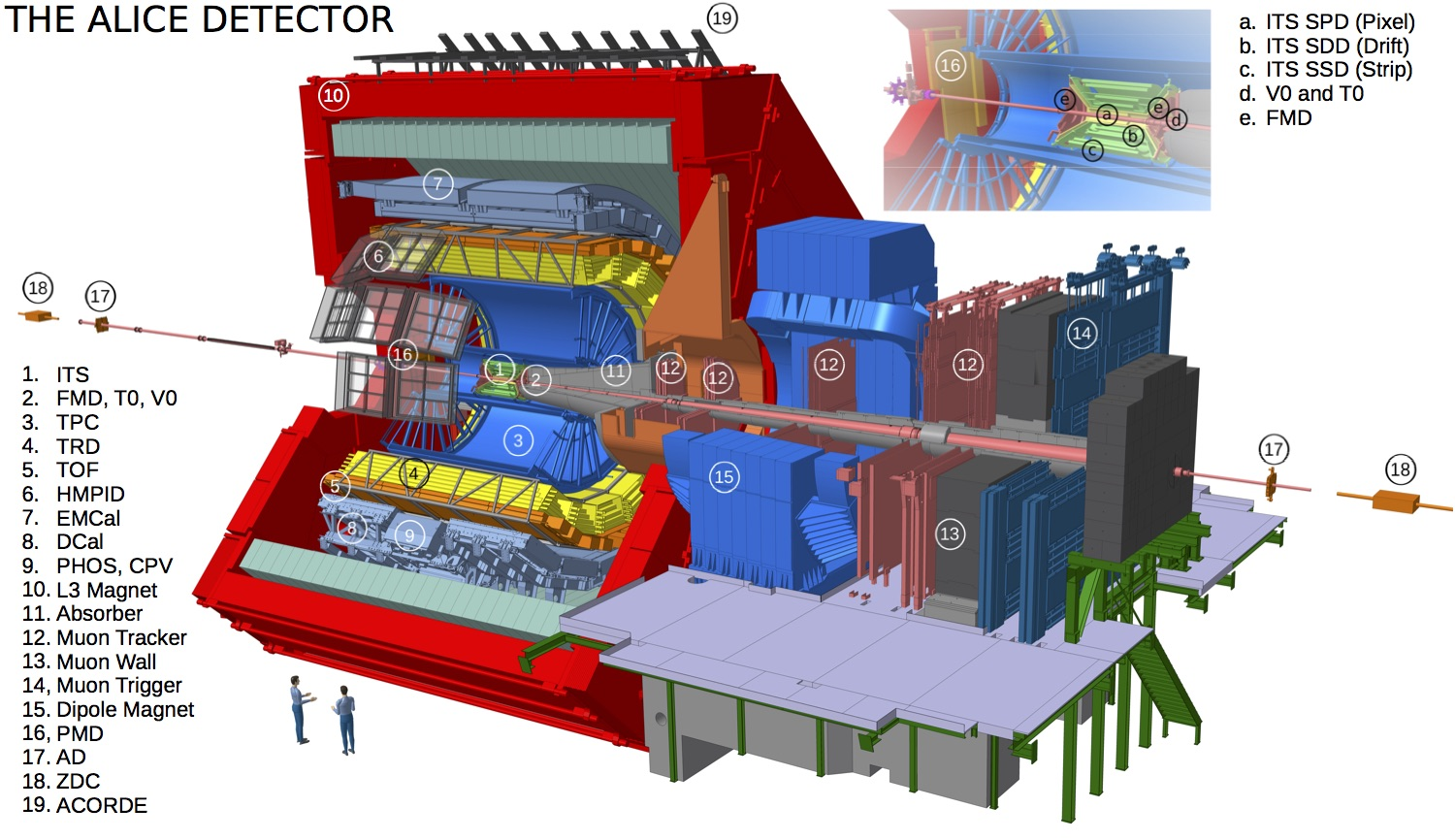}
    \caption{Components of the~ALICE detector in its Run 2 configuration \cite{botta:2017a}.}
    \label{fig:alice-detectors}
\end{figure}

The ALICE experiment is composed of several sub-detectors, some of which measure particle properties that can be used for identification. Figure \ref{fig:alice-detectors} presents a scheme of the detector in Run 1 and Run 2 LHC data-taking periods.

The three detectors particularly useful for PID are: Time Projection Chamber (TPC), Time-of-Flight (TOF), Transition Radiation Detector (TRD). TPC is one of the most important ALICE detectors as it records 3D information of the~trajectory of the~particles, and particle ionization energy loss essential for particle identification. The Time-of-Flight detector measures  particle travel time from the collision vertex to the detector, from which particle velocity and mass are calculated. TRD records transition radiation, that is, the emission of a few photons by traversing electrons, which helps in distinguishing electrons from other charged particles.

With the signals recorded by the detectors described above, particles are chosen using selection criteria compared with theoretical calculations. The traditional method compares the number of standard deviations from the expected value for all detector signals. Particles falling outside the selection region are rejected. However, when the characteristics of different particle species overlap, combining information from multiple detectors becomes difficult. Choosing selection regions by trial and error is less effective in this case, lowering PID efficiency and limiting the statistical significance of the final analysis.

These shortcomings can be addressed with more advanced classification methods such as Bayesian models or neural-network-based (NN) approaches. The Bayesian approach~\cite{ALICE:2016zzl} is available in the new ALICE software framework O$^2$, but its flexibility is limited.
For these reasons, we studied possible improvements employing neural networks for PID.

\section{Machine learning for PID}
\label{sec:ml-for-pid}

Particle identification is a standard classification problem for a machine learning algorithm. The inputs to the model are various track parameters, and from them, machine learning (ML) can utilize more information and learn more complex relationships than what a human analyzer can observe. Thanks to the present popularity of neural networks, there are many software libraries that enable fast implementation of a given neural network model.

We start with a simple feed-forward neural network, which is trained and applied on Monte Carlo simulated data. The network acts as a binary classifier, outputting predictions for a given particle species. The single-value output is normalized to the range $(0, 1)$ by applying the logistic function $f(x) = \frac{1}{1+e^{-x}}$. The classifier result approximates the probability of the example corresponding to a specific particle type. Additionally, the user can decide which detector information they want to use. This has an impact on the size of the input vector, and requires the development of different models for different input sizes. The networks for each (anti)particle species and detector setup are trained independently. Initial results are promising: a simple network using combined information from TPC and TOF detectors can improve efficiency compared to traditional method while not losing purity.

One must also consider the limitations that an ML model will have. Firstly, the performance greatly depends on training data quality, so also on the results of algorithms used for simulation, detector calibration, and track reconstruction. Physics studies are usually performed on well-calibrated, high-quality data, which further predisposes the use of ML. However, by design, such data still contains many incomplete samples. The problem and our solution are described in detail in Section~\ref{sec:incomplete}. Another particularity is the domain shift between the distributions of variables in a training dataset (simulation) and a corresponding test/inference dataset (real experiment data). This is mitigated with the domain adaptation technique explained in Section~\ref{domain-adaptation}.

Next, it is more difficult to obtain systematic uncertainties from ML results. A few approximate methods exist. For example, in~\cite{ghosh2021uncertainty}, the authors propose a model where classifiers are aware of systematic uncertainties of input parameters. This approach improves the model's sensitivity. In~\cite{englert2019machine}, the authors use adversarial training for this purpose.

\subsection{Integration of PID ML with the O$^2$ framework}
\label{sec:o2-pid}

Applying machine learning models in an analysis brings the technical challenge of crossing different programming languages. Various ML frameworks used in ALICE are implemented in Python, PID ML not being an exception. However, the ALICE analysis framework, O$^2$~\cite{o2physics}, is written in C++ and heavily utilizes the ROOT~\cite{BRUN199781} library. Our solution makes use of the ONNX (Open Neural Network Exchange) standard~\cite{onnx}, which defines a common file format for storing machine learning models developed in various frameworks such as Tensorflow~\cite{tensorflow2015-whitepaper} and PyTorch~\cite{pytorch}. The ONNXRuntime~\cite{onnxruntime} can then be used for ONNX model inference in C++.

\begin{figure}
    \centering
    \includegraphics[width = \textwidth]{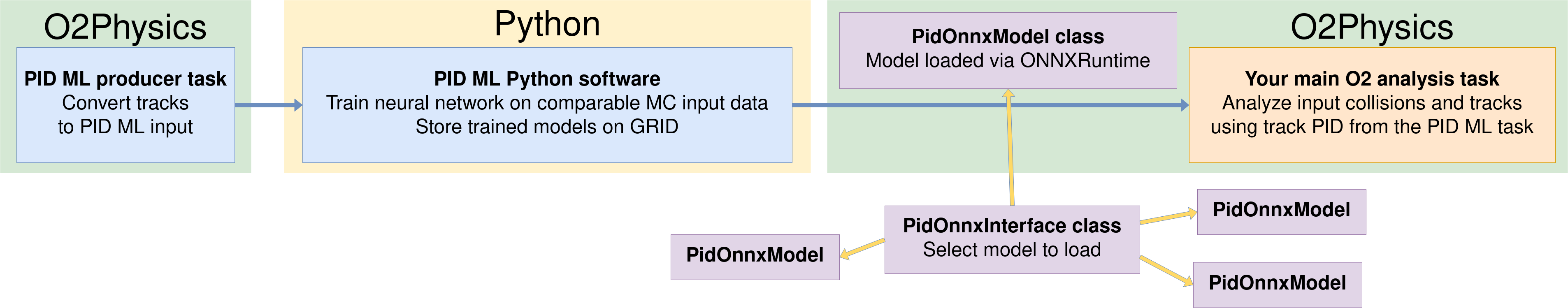}
    \caption{Scheme of the current ML particle identification workflow in O$^2$. More information about the workflow topology in the O$^2$ Analysis Framework can be found in~\cite{Alkin:2021mfo}.}
    \label{fig:pidml-interface}
\end{figure}

Figure~\ref{fig:pidml-interface} depicts the current workflow of PID ML in O$^2$. The input data consists of reconstructed collisions and tracks in AO2D files, which are stored on disk in ROOT file format. A PIDML producer task takes the initial AO2D files, applies a few rough preselections to exclude meaningless tracks, and produces skimmed data that contains only track properties used by a neural network. The data is then used for model training. Trained models are stored in the ONNX format in the Condition and Calibration Data Base (CCDB). The database is accessible by analyses running on the worldwide LHC computing GRID. An analyzer can apply an ML model for particle identification by adding an instance of the PidOnnxModel class or, in more complex use cases, the PidOnnxModelInterface class. The latter enables handy management of multiple ML models, each with different target particle species, detector setup, and acceptance threshold.

The same solution, with ONNXRuntime and a C++ class to wrap management of a single ONNX model, was applied in other machine learning projects in O$^2$. A common interface that would further unify various implementations of the ONNX integration is being developed based on PidOnnxModel, and it is already used for supporting calculations for TPC PID and selection of $\Lambda_c^+$ in the $\rm{pK}^-\pi^+$ decay channel.

\section{Dealing with incomplete data}
\label{sec:incomplete}

The number of inputs to the network can vary not only because of user preferences but also at random, because of independent measurements of particle properties by various detectors. A particle may be measured by a subset of detectors while not being recorded by others. For instance, a detector may malfunction or be switched off, or the particle can have characteristics (i.e., too low transverse momentum) that do not match the detector specification. 

A detailed study was performed to find the best model for particle identification with missing data. The simplest method is case deletion, that is, removing incomplete samples from data. However, it prevents the identification of samples with missing detector information, which are equally important as the samples with a complete set of signals. On the other hand, imputation techniques work by filling missing data with artificial values computed as a mean or median of existing signals, or with values predicted with a simple model such as linear regression. Both case deletion and imputation distort the initial dataset, which can disturb predictions of the neural network.

Instead of changing data, one can adjust the model architecture. A straightforward solution is to use a neural network ensemble: a set of classifiers, one per each subset of the training dataset without missing data. In particular, in neural network reduction~\cite{sharpe1995dealing}, authors propose to split the dataset into the largest possible complete subsets. Nevertheless, this method can be computationally expensive, especially with a growing set of attributes with missing values.

\begin{figure}
    \centering
    \includegraphics[width=\textwidth]{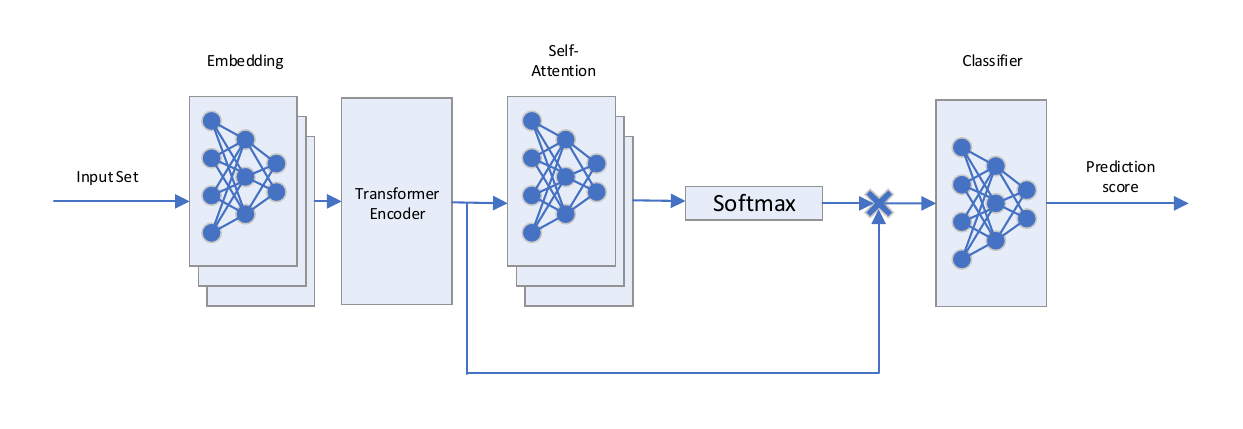}
    \caption{The proposed model architecture. Layered blocks are applied separately to each vector in a set. Single blocks are applied to their input as a whole.} \label{fig:model_arch}
\end{figure}

To overcome the aforementioned shortcomings, we introduce a novel method based on the attention mechanism, similar to the method introduced for a medical use-case in AMI-Net~\cite{wang2019attention}. The overview of the system is shown in Figure~\ref{fig:model_arch}.

First, all data examples are encoded into a set of feature-value pairs~\cite{grangier2010feature}, independently of their characteristics regarding missing data. Each pair consists of a non-missing value in the input data and the index of that value with one-hot encoding applied. Then, the feature-value pairs are transformed by a neural network with a single hidden layer into embeddings that place similar features close in the embedded space.

The Transformer's~\cite{vaswani2017attention} encoding module is applied to the set of embedding vectors to connect the different features each vector represents and find input patterns. For example, a measurement from a specific detector could be used if the momentum is within a particular range. The softmax function is applied to the attention output, a variable-size set of vectors. An additional self-attention layer is used to merge these vectors into a single one.

Finally, the pooled vector is processed by the simple classifier described in Section~\ref{sec:ml-for-pid}.

\subsection{Results}
\label{sec:missing-results}

The data comes from a Monte Carlo simulation of proton--proton collisions at $\sqrt{s}=13$~TeV with realistic simulation of the time evolution of the detector conditions in the LHC Run 2 data-taking period. The simulation was performed with Pythia8~\cite{Sjostrand:2014zea}, the Geant~4~\cite{Brun:1994aa} particle transport model, and general-purpose settings.

\begin{figure}
    \centering
    \resizebox{0.5\textwidth}{!}{
        \begin{tabular}[b]{ccccc}
\toprule
pion & kaon & proton & electron & muon \\
\midrule
43.59\% & 3.415\% & 2.026\% & 0.794\% & 0.323\% \\
\cline{1-5}
\bottomrule
\toprule
antipion & antikaon & antiproton & antielectron & antimuon \\
\midrule
43.66\% & 3.288\% & 1.819\% & 0.762\% & 0.321\% \\
\cline{1-5}
\bottomrule
\end{tabular}
    }
    \hfill
    \includegraphics[width=0.45\textwidth]{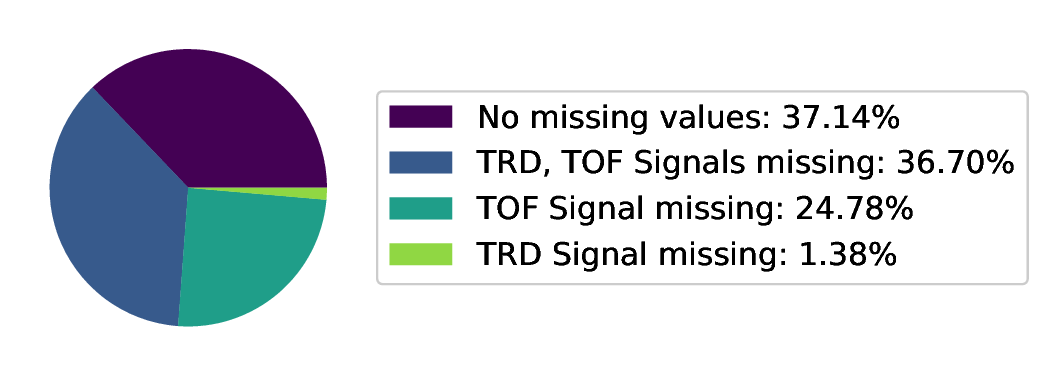}
    \caption{Left: Particle type distribution in studied dataset. Approximately 97.8\% of the examples belong to the 6 most populous particle types. Right: Missing data distribution. Over 62.8\% of the examples is missing at least one value.}
    \label{fig:part_val_distr}
\end{figure}

The dataset consists of 2 751 934 examples with track's transverse momentum $\pt \geq 0.1$ \GeVc. 95\% of examples fall into the range $[0.12, 1.76]$ \GeVc. Each example contains 19 features, representing signals from TPC, TRD, and TOF detectors as well as predetermined properties such as spatial coordinates of a track reconstruction starting point ($x$, $y$, $z$, angle $\alpha$), track momentum, type, charge, and the distances of closest approach (DCA) of the track trajectory to the collision's primary vertex measured in the $xy$ plane ($d_{xy}$) and the $z$ direction ($d_{z}$). The examples are labeled with ten different particle types. The distribution of particles and their missing values is shown in Figure \ref{fig:part_val_distr}.

For each compared method, we train identical binary classification models for 6 of the most abundant particle types: pions, protons, kaons, and their respective antiparticles. All trained models are identical for imputation methods and the neural network ensemble. They have three hidden layers of sizes 64, 32 and 16, and a single output.
Between the layers, we use Rectified Linear Unit (ReLU) activation function $f(x) = \max({0, x})$. Dropout regularization with a rate of 0.1 is applied after each activation layer. For imputation techniques, all models process inputs of size 19, as all missing features are imputed. For the neural net ensemble, the four networks have input sizes 19, 17, 17, and 15, depending on the combination of missing values from the TRD and TOF detectors (2 features per each detector).
The tested attention architecture described in Section~\ref{sec:incomplete} is implemented with ReLU activation between neural network layers, and dropout regularization with a rate of 0.1 at the output of the embedding layer and the output of each sub-layer in the encoder.
Model parameters are selected using a hyperparameter sweep for a fair comparison of different network architectures.

Each model is trained using all complete and incomplete examples and tested in two cases: (1) with all available test examples and (2) with the complete ones only. For case deletion, no results are available for testing on the missing data dataset. The models are evaluated on test-set with standard metrics -- precision (purity) and recall (efficiency). Since precision and recall are antagonistic, also $F_1$ metric is considered, which is a combination of the first two: $\rm{F}_1=\frac{precision~\cdot~recall}{precision~+~recall}$. The scored means and standard deviations of 4 runs with different seeds for classification of 3 particle species are shown in Table~\ref{tab:results_table}.  Analogous results are obtained for the respective antiparticle species. The best-performing algorithm is highlighted in green, and the second best -- in blue.

\begin{table}[b]
    \def\arraystretch{1.3}
    \caption{Classification results for the three most common particle species.}
    \label{tab:results_table}
   
    \resizebox{0.32\textwidth}{!}{
        \begin{tabular}{lccc}
\toprule
\multicolumn{4}{c}{$\pi$ (pions), only complete data} \\
model & precision & recall & $F_1$ \\
\midrule
Delete & 0.9908 $\pm$ 0.0007 & \cellcolor{blue-rec}0.9967 $\pm$ 0.0004 & \cellcolor{blue-rec}0.9937 $\pm$ 0.0001 \\
\cline{1-4}
Ensemble & \cellcolor{green-rec}0.9911 $\pm$ 0.0004 & 0.9964 $\pm$ 0.0006 & \cellcolor{green-rec}0.9938 $\pm$ 0.0001 \\
\cline{1-4}
Mean & 0.9885 $\pm$ 0.0009 & \cellcolor{green-rec}0.9969 $\pm$ 0.0004 & 0.9927 $\pm$ 0.0004 \\
\cline{1-4}
Proposed & \cellcolor{blue-rec}0.9908 $\pm$ 0.0002 & 0.9964 $\pm$ 0.0003 & 0.9936 $\pm$ 0.0001 \\
\cline{1-4}
Regression & 0.9902 $\pm$ 0.0002 & 0.9949 $\pm$ 0.0014 & 0.9925 $\pm$ 0.0007 \\
\cline{1-4}
\bottomrule
\end{tabular}

    }
    \hfill
    \resizebox{0.32\textwidth}{!}{
        \begin{tabular}{lccc}
\toprule
\multicolumn{4}{c}{p (protons), only complete data} \\
model & precision & recall & $F_1$ \\
\midrule
Delete & \cellcolor{blue-rec}0.9923 $\pm$ 0.0032 & 0.9963 $\pm$ 0.0005 & 0.9943 $\pm$ 0.0016 \\
\cline{1-4}
Ensemble & 0.9916 $\pm$ 0.0024 & \cellcolor{green-rec}0.9976 $\pm$ 0.0007 & 0.9946 $\pm$ 0.0013 \\
\cline{1-4}
Mean & 0.9922 $\pm$ 0.0019 & \cellcolor{blue-rec}0.9972 $\pm$ 0.0004 & \cellcolor{blue-rec}0.9947 $\pm$ 0.0008 \\
\cline{1-4}
Proposed & \cellcolor{green-rec}0.9928 $\pm$ 0.0010 & 0.9968 $\pm$ 0.0009 & \cellcolor{green-rec}0.9948 $\pm$ 0.0002 \\
\cline{1-4}
Regression & 0.9910 $\pm$ 0.0009 & 0.9965 $\pm$ 0.0009 & 0.9937 $\pm$ 0.0007 \\
\cline{1-4}
\bottomrule
\end{tabular}

    }
    \hfill
    \resizebox{0.32\textwidth}{!}{
        \begin{tabular}{lccc}
\toprule
\multicolumn{4}{c}{K (kaons), only complete data} \\
model & precision & recall & $F_1$ \\
\midrule
Delete & \cellcolor{green-rec}0.9693 $\pm$ 0.0037 & 0.9698 $\pm$ 0.0026 & 0.9695 $\pm$ 0.0006 \\
\cline{1-4}
Ensemble & 0.9665 $\pm$ 0.0038 & \cellcolor{blue-rec}0.9782 $\pm$ 0.0031 & \cellcolor{green-rec}0.9723 $\pm$ 0.0010 \\
\cline{1-4}
Mean & \cellcolor{blue-rec}0.9683 $\pm$ 0.0017 & 0.9533 $\pm$ 0.0067 & 0.9608 $\pm$ 0.0036 \\
\cline{1-4}
Proposed & 0.9603 $\pm$ 0.0098 & \cellcolor{green-rec}0.9806 $\pm$ 0.0072 & \cellcolor{blue-rec}0.9704 $\pm$ 0.0017 \\
\cline{1-4}
Regression & 0.9427 $\pm$ 0.0098 & 0.9701 $\pm$ 0.0051 & 0.9562 $\pm$ 0.0039 \\
\cline{1-4}
\bottomrule
\end{tabular}

    }
    \vspace{5mm}

    \resizebox{0.32\textwidth}{!}{
        \begin{tabular}{lccc}
\toprule
\multicolumn{4}{c}{$\pi$ (pions), all data} \\
model & precision & recall & $F_1$ \\
\midrule
Ensemble & \cellcolor{blue-rec}0.9747 $\pm$ 0.0025 & 0.9946 $\pm$ 0.0021 & \cellcolor{blue-rec}0.9845 $\pm$ 0.0004 \\
\cline{1-4}
Mean & 0.9731 $\pm$ 0.0007 & \cellcolor{blue-rec}0.9952 $\pm$ 0.0007 & 0.9840 $\pm$ 0.0001 \\
\cline{1-4}
Proposed & \cellcolor{green-rec}0.9749 $\pm$ 0.0006 & \cellcolor{green-rec}0.9954 $\pm$ 0.0005 & \cellcolor{green-rec}0.9850 $\pm$ 0.0002 \\
\cline{1-4}
Regression & 0.9733 $\pm$ 0.0006 & 0.9949 $\pm$ 0.0007 & 0.9840 $\pm$ 0.0004 \\
\cline{1-4}
\bottomrule
\end{tabular}

    }
    \hfill
    \resizebox{0.32\textwidth}{!}{
        \begin{tabular}{lccc}
\toprule
\multicolumn{4}{c}{p (protons), all data} \\
model & precision & recall & $F_1$ \\
\midrule
Ensemble & 0.9716 $\pm$ 0.0046 & \cellcolor{blue-rec}0.9374 $\pm$ 0.0030 & 0.9542 $\pm$ 0.0012 \\
\cline{1-4}
Mean & \cellcolor{green-rec}0.9785 $\pm$ 0.0041 & 0.9334 $\pm$ 0.0032 & \cellcolor{blue-rec}0.9554 $\pm$ 0.0006 \\
\cline{1-4}
Proposed & \cellcolor{blue-rec}0.9780 $\pm$ 0.0044 & \cellcolor{green-rec}0.9386 $\pm$ 0.0027 & \cellcolor{green-rec}0.9579 $\pm$ 0.0007 \\
\cline{1-4}
Regression & 0.9738 $\pm$ 0.0040 & 0.9367 $\pm$ 0.0038 & 0.9549 $\pm$ 0.0015 \\
\cline{1-4}
\bottomrule
\end{tabular}

     }
    \hfill
    \resizebox{0.32\textwidth}{!}{
        \begin{tabular}{lccc}
\toprule
\multicolumn{4}{c}{K (kaons), all data} \\
model & precision & recall & $F_1$ \\
\midrule
Ensemble & \cellcolor{blue-rec}0.9118 $\pm$ 0.0200 & \cellcolor{blue-rec}0.8272 $\pm$ 0.0142 & \cellcolor{blue-rec}0.8674 $\pm$ 0.0016 \\
\cline{1-4}
Mean & 0.9083 $\pm$ 0.0171 & 0.8232 $\pm$ 0.0096 & 0.8636 $\pm$ 0.0034 \\
\cline{1-4}
Proposed & \cellcolor{green-rec}0.9155 $\pm$ 0.0071 & \cellcolor{green-rec}0.8368 $\pm$ 0.0082 & \cellcolor{green-rec}0.8744 $\pm$ 0.0014 \\
\cline{1-4}
Regression & 0.9117 $\pm$ 0.0100 & 0.8178 $\pm$ 0.0021 & 0.8622 $\pm$ 0.0046 \\
\cline{1-4}
\bottomrule
\end{tabular}

    }
\end{table}

It is clear that the proposed attention architecture achieves very good scores of each metrics, comparable with other analyzed models. It performs particularly well with the realistic dataset containing incomplete samples. At the same time, our model avoids the flaws of other solutions: artificial bias in imputed and case-deleted data, and potentially larger complexity of neural network ensemble.

\section{Domain adaptation}
\label{domain-adaptation}

In analysis, PID is used to select particles of desired types in both real experimental data and Monte Carlo simulations. However, recorded signals in physical detectors can differ from those produced in simulations. Standard PID methods rely on partially automated processes for data domains alignment, e.g., ALICE uses a tuning of simulated signals in such a way they reproduce, on average, the collected data distributions of specific signals.

To circumvent the limitations of standard domain alignment methods, we propose combining domain alignment with particle identification stating it as a known problem of classification with unsupervised domain adaptation. The main idea of this technique is to learn the discrepancies between two data domains, that is, the labeled source domain (simulation data) and the unlabeled target one (experimental data), and translate those to a single hyperspace. Classifiers trained on top of features located in combined latent space should have similar performance on both MC simulated and experimental data.

\begin{figure}[t]
    \centering
    \includegraphics[width=0.6\textwidth]{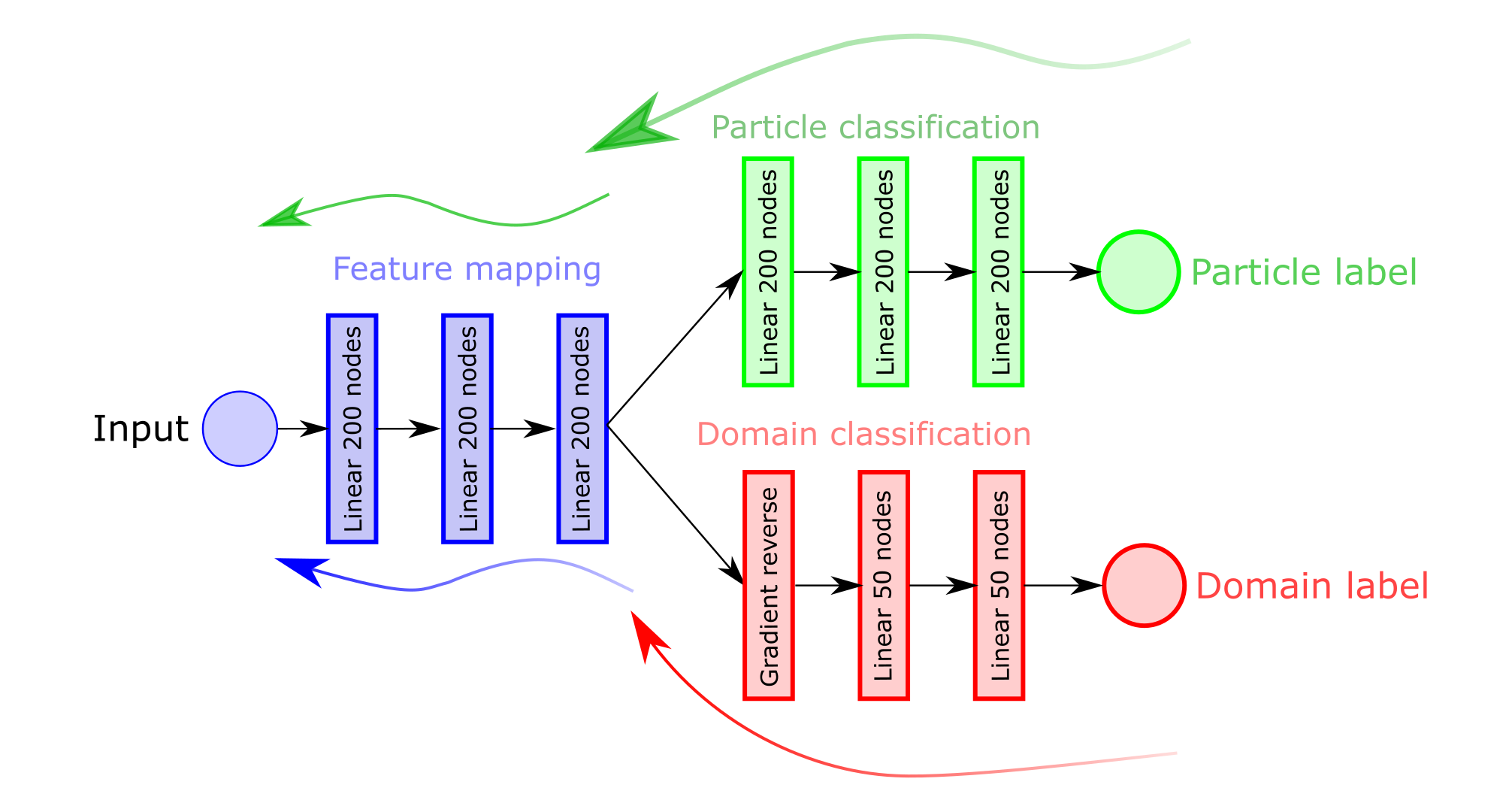}
    \caption{Architecture of Domain Adversarial Neural Network.}
    \label{fig:pid-dann}
\end{figure}

A Domain Adversarial Neural Network (DANN) applies domain adaptation in neural networks. The architecture, presented in Figure~\ref{fig:pid-dann}, is composed of three neural networks. The goal of the feature mapping network is to map original input into domain invariant features. Those features serve as an input to the standard particle classifier that outputs the particle type. Additionally, the last model, known as domain classifier, enforces domain invariance of extracted features through adversarial training. The training of the model is divided into two steps. First, on top of current features from the feature mapping network, the domain classifier is trained independently to classify domain labels -- whether data come from a real or a simulation source. Then, the domain classifier is kept frozen so that the particle classifier and the feature mapper can be trained jointly to predict accurate particle types while fouling the domain classifier. With this approach, the feature mapper's weights are updated with a gradient from the particle classifier and reversed gradient from the domain classifier. Training a domain-adaptation-based classifier is more complex than the classical neural model. However, the application performance of those two methods is similar and depends on the complexity of the classifier and feature extractor.

The domain-adaptation technique is widely used in natural language processing~\cite{blitzer2007biographies,glorot2011domain} and computer vision~\cite{gopalan2011domain,fernando2013unsupervised}. In the domain of high-energy physics, the author of~\cite{walter2018domain} presents how this method can improve the quality of automatic jet tagging on real experimental data. The initial tests with Domain Adversarial Neural Networks (DANN)~\cite{ganin2016domain} show that this technique improves the classification of particles in experimental data.

\section{Future developments and conclusions}
\label{conclusions}

The current priority is to test the framework in a real-world analysis on simulation data from the present Run 3 data-taking period. Afterward, we will extend the attention model with domain adaptation and test it on the new real data. Finally, we will be able to achieve regular production of models for Run 3. 

In order to estimate statistical uncertainties, we plan to include tests analogous to~\cite{jha2019impact} measuring how training dataset selection might affect the model's performance. We intend to follow the developments described in~\cite{gal2016dropout}, where the authors show that dropout, a standard method for reducing overfitting in neural networks, might approximate Bayesian uncertainty in deep Gaussian processes. A method for calibrating Bayesian uncertainties is also presented in~\cite{kuleshov2018accurate}.

\section*{Acknowledgments}
The research was funded by POB HEP of Warsaw University of Technology within the Excellence Initiative: Research University (IDUB) program, by the Polish Ministry of Education and Science in the framework of International Co-financed Projects under agreements no. 2022/WK/01, 5236/CERN/2022/0 and by National Science Centre under agreement UMO-2021/43/D/ST2/02214.

\bibliography{bibliography}

\end{document}